\documentclass[twocolumn]{aastex631}

\begin{document}

\title{Binary Disruption and Ejected Stars from Hierarchical Star Cluster Assembly}

\correspondingauthor{Claude Cournoyer-Cloutier}
\email{cournoyc@mcmaster.ca}

\author[0000-0002-6116-1014]{Claude Cournoyer-Cloutier}\thanks{Claude Cournoyer-Cloutier and Jeremy Karam are co-first authors}
\affiliation{Department of Physics and Astronomy, McMaster University, 1280 Main Street West, Hamilton, ON, L8S 4M1, Canada}

\author[0000-0003-3328-329X]{Jeremy Karam}\thanks{Claude Cournoyer-Cloutier and Jeremy Karam are co-first authors}
\affiliation{Department of Physics and Astronomy, McMaster University, 1280 Main Street West, Hamilton, ON, L8S 4M1, Canada}

\author[0000-0003-3551-5090]{Alison Sills}
\affiliation{Department of Physics and Astronomy, McMaster University, 1280 Main Street West, Hamilton, ON, L8S 4M1, Canada}

\author[0000-0001-5839-0302]{Simon Portegies Zwart}
\affiliation{Sterrewacht Leiden, Leiden University, Niels Bohrweg 2, 2333 Leiden, The Netherlands}

\author[0000-0002-3001-9461]{Maite J.~C. Wilhelm}
\affiliation{Sterrewacht Leiden, Leiden University, Niels Bohrweg 2, 2333 Leiden, The Netherlands}



\begin{abstract}
We simulate mergers between star clusters embedded within their natal giant molecular cloud. We extract initial conditions from cloud-scale simulations of cluster formation and introduce different prescriptions for primordial binaries.
We find that simulations that do not include primordial binaries result in a larger fraction of unbound stars than simulations which include a prescription for binaries based on observations. We also find a preferred direction of motion for stars that become unbound during the merger. 
Sub-cluster mergers within realistic gas environments promote binary disruption while mergers between idealized, gas-rich spherical clusters do not produce the same disruption.
Binary systems with smaller semi-major axes are disrupted in simulations of sub-cluster mergers within their natal environment compared to simulations that do not include the realistic gas environment. We conclude that binary disruption and the production of an anisotropic distribution of unbound stars are the natural consequences of sub-cluster mergers during star cluster assembly. 
\end{abstract}



\section{Introduction} \label{sec:intro}
Star clusters form hierarchically within giant molecular clouds~\citep[GMCs, e.g.][]{Lada2003, PortegiesZwart2010}. Simulations~\citep[e.g.][]{Dobbs2022, Fujii2012, Fujii2022, Grudic2018, Howard2018, Rieder2022} have shown that massive star clusters assemble through the merger of smaller sub-clusters. Observations also indicate that Orion~\citep{Fujii2022}, Westerlund~2~\citep{Sabbi2012, Zeidler2021} and R136~\citep{Fahrion2024, Fujii2012}, among others, are assembling hierarchically. Simulations further suggest that sub-cluster mergers result in clusters that better match the density profiles~\citep{Fujii2012} of observed young massive clusters than isolated cluster formation, as well as providing a natural explanation for the light element abundance variations~\citep{Howard2019, Lahen2024} observed in globular clusters. Sub-cluster mergers are also dynamically rich processes, that can enhance the formation of new stars from compression of the surrounding gas~\citep{Fujii2022} and impart dynamical signatures in the stellar component of the clusters involved in the merger ~\citep[e.g.][]{Fujii2022, Karam2024}.
Studying mergers between embedded star clusters therefore allows us to gain physical insights about the process of massive cluster formation.

Most stars and protostars in star-forming regions are part of a binary (or higher-order) system, with a multiplicity fraction increasing strongly with stellar mass~\citep[see][for recent reviews]{Moe2017, Offner2023}. The interplay between stellar multiplicity and stellar clustering is non-trivial. Observations reveal that -- for solar-mass stars, at least -- stellar multiplicity in young clusters depends on cluster density: higher-density environments~\citep[e.g. Orion,][]{Duchene2018} have fewer wide binaries than low-density environments~\citep[e.g. Taurus,][]{Kraus2011}. 
Recent simulations of embedded~\citep{Cournoyer-Cloutier2021} and gas-free~\citep{Torniamenti2021} cluster assembly further suggest that changes to populations of binaries take place while the cluster is forming but that binary properties are then stable following gas expulsion. 
Similar trends with environment are observed in older clusters. \citet{Deacon2020} compare binaries in moving groups and open clusters, and find that stars formed in looser associations have more wide companions than stars formed in cluster-forming regions. 
For massive, dense globular clusters, stellar multiplicity is anti-correlated with cluster luminosity~\citep[and implicitly cluster mass, ][]{Milone2016}, but not with central density. Taken together, those observations suggest that the differences in the populations of binaries are a result of a cluster's assembly process, rather than only its present-day properties.

Accounting  for the high multiplicity fraction of O and B stars, which are short-lived and rare, is simultaneously both more complicated and more important. Massive OB stars regulate the subsequent star formation in their natal clouds via feedback: they produce winds, radiation, and supernova explosions that heat the nearby gas and drive it away from the central, star-forming
regions, thus preventing the formation of other stars within the cluster. 
Massive stars routinely escape from their natal cluster; for example, recent observations show OB-stars moving away from the young massive clusters M16~\citep{Stoop2023}, M17~\citep{Stoop2024}, and NGC 3603~\citep{Kalari2019}. The removal of massive stars from their birth cluster can have consequences on galactic scales, due to the enrichment and supernova feedback from stars into the interstellar medium~\citep{Andersson2020}.

Runaway stars are often classified as stars with radial velocities $v_r \gtrsim 30$ km/s. They can be produced through few-body dynamical interactions in a cluster, involving at least one binary~\citep[e.g.][]{Podeva1967, Hoogerwerf2000, Fujii2011, Gvarmadze2011}. Through the interaction, the binary becomes more tightly bound and the energy lost is transferred to another star(s) as kinetic energy; if the amount of energy is large enough, the star's velocity may exceed the cluster's escape velocity and the star may become unbound from the cluster. 
Two populations of OB runaways are observed in 30 Doradus~\citep{Sana2022}, attributed to binary disruption following a supernova~\citep[originally proposed by][]{Blaauw1961} and to few-body encounters within a star cluster. Simulations have shown that merger between sub-clusters may also enhance the production of runaway stars~\citep{Fujii2022, Polak2024}. These simulations however did not include primordial binaries, and so could not compare the relative effects of sub-cluster mergers and few-body encounters on the production of runaway stars during cluster assembly.

Any understanding of star cluster formation therefore requires an understanding of how a population of massive binaries evolves during cluster formation. Massive OB binaries are formed within embedded clusters, and the population of massive binaries in clusters is less dynamically processed than that of massive stars in the field, which are often runaways. There is however growing evidence supporting significant hardening of massive binaries during the process of cluster formation~\citep[e.g.][]{Sana2017, Ramirez-Tannus2021, Bordier2022}. It is clear that there is some highly non-trivial interplay between cluster mergers, binaries, and runaway stars. Some previous studies have started exploring this interplay~\citep[e.g.][]{Fujii2012}, using idealized initial conditions. 

In this work, we utilize initial conditions drawn from a cluster formation simulation~\citep[][Wilhelm et al. in prep.]{Wilhelm2024}\footnote{\url{https://hdl.handle.net/1887/3717680}} that includes a sub-cluster merger resulting in the production of runaway stars. The simulation only contains two sub-clusters, allowing us to cleanly isolate the merger. It was performed with \textsc{Torch}~\citep{Wall2019, Wall2020} and includes star formation and stellar feedback, but does not however have any population of primordial binaries\footnote{A version of \textsc{Torch} does include binaries~\citep{Cournoyer-Cloutier2021}, however it was not used for the simulations from which we extract our initial conditions.}. We introduce primordial binaries to those initial conditions, and run a suite of simulations with different prescriptions for binaries. We run our simulations using only stellar dynamics and hydrodynamics. Those computational methods that are simpler and less expensive than those used in the \textsc{Torch} framework, allowing us to explore a larger range of parameters. In this paper, we study how sub-cluster mergers within giant molecular clouds influence a population of binary stars, and study the production of unbound stars during sub-cluster mergers. In Section~\ref{sec:methods}, we describe our simulation methods and initial conditions. We present the results in Section~\ref{sec:results} and discuss their implications for hierarchical cluster formation in Section~\ref{sec:discussion}. We summarize our key conclusions in Section~\ref{sec:summary}. 

\section{Methods} \label{sec:methods}
\begin{figure}[tb!]
    \centering
    \includegraphics[width=\linewidth, clip=True, trim=1.5cm 1.2cm 1cm 0.5cm]{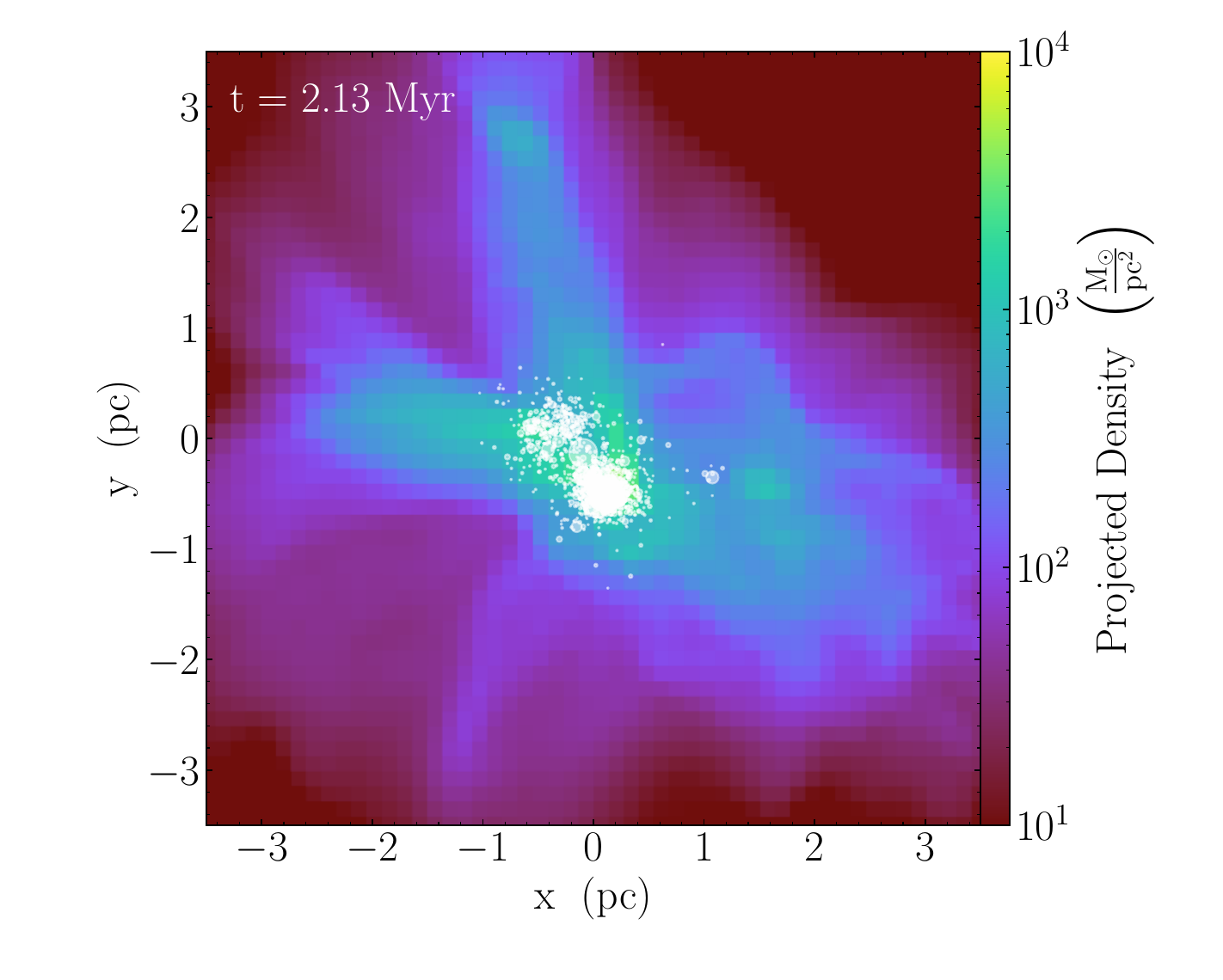}
    \caption{Initial conditions from the full \textsc{Torch} run, zoomed in to show the merging clusters. The full domain width is 17.5 pc, and the central 7 pc are shown. The surface density shown is calculated using the resolution of the gas cells, 0.136 pc.}
    \label{fig:Torch ICs}
\end{figure}

We use initial conditions extracted from a simulation of star formation within a collapsing cloud with an initial mass of $10^4$ M$_{\odot}$, which naturally gives rises to a merger between two sub-clusters of stars with stellar masses of 1684.3 and 239.5 M$_{\odot}$ in 2603 and 444 stars. The merger takes place 2.13 Myr after the start of the simulation, and 0.66 Myr after the formation of the first star. The snapshot we use as our initial conditions is shown in Figure~\ref{fig:Torch ICs}. In Section~\ref{sec:Torch}, we list the physics included in the full simulation. In Sections~\ref{sec:hydro} and~\ref{sec:Petar}, we describe the numerical methods used in our simulations, while the initial conditions are outlined in Section~\ref{sec:ICs}. 

\subsection{\textsc{Torch} simulation} \label{sec:Torch}

The simulation from which we draw our initial conditions was done using \textsc{Torch}~\citep{Wall2019, Wall2020} and is presented in detail in~\citet{Wilhelm2024}. In the text below, we describe the relevant physics implemented in the \textsc{Torch} simulation, and highlight how they lead to realistic initial conditions for the simulations presented in this work.

The \textsc{Torch} simulation uses magneto-hydrodynamics (MHD) coupled with stellar dynamics (including a treatment of dynamically-formed binaries), star formation, stellar evolution and feedback, and a sub-grid model for protoplanetary disks coupled through the \textsc{Amuse} framework~\citep{PortegiesZwart2009, Pelupessy2013, PortegiesZwart2013, AMUSEbook}. 
MHD is handled by the adaptive mesh refinement code \textsc{Flash}~\citep{Fryxell2000, Dubey2014}, with a maximum spatial resolution of 0.136 pc. In the \textsc{Torch} simulation, star formation takes place in bound regions with high density gas ($\geqslant 3.819$ x 10$^{-21}$ g cm$^{-3}$) and converging flows. This results in the formation of sub-clusters with shapes consistent with observed embedded clusters~\citep[][]{Cournoyer-Cloutier2023}. The stars are sampled from a~\citet{Kroupa2001} initial mass function (IMF) from 0.08~M$_{\odot}$ to 100~M$_{\odot}$. 
Stars more massive than 7 M$_{\odot}$ provide feedback to the simulation in the form of momentum-driven winds, ionizing radiation and radiation pressure~\citep[][]{Wall2020}, resulting in a non-smooth distribution of gas as seen in real star-forming regions.

\subsection{N-body dynamics} \label{sec:Petar}
We  use the stellar masses, positions and velocities from the full \textsc{Torch} run directly. We handle stellar dynamics with \textsc{Petar}~\citep{Wang2020b}, which is optimized to handle large numbers of binaries and few-body encounters. \textsc{Petar} handles long-range interactions with a Barnes-Hut tree~\citep[][as implemented by~\citealt{Iwasawa2016}]{Barnes1986}, short-range interactions with a fourth-order Hermite integrator~\citep{Makino1992}, and stable binaries and close encounters with the slow-down algorithmic regularization method~\citep[][\textsc{sdar}]{Wang2020a}. The longest timestep dt$_{\mathrm{soft}}$ of the simulation, used for long-range interactions, must be chosen in conjunction with the changeover radii r$_{\mathrm{in}}$ and r$_{\mathrm{out}}$ between short- and long-range interactions. 
We set dt$_{\mathrm{soft}}$ to 1/100$^{\mathrm{th}}$ of the shortest orbital period we could have for a binary in the long-range interaction regime.
We set r$_{\mathrm{out}} = 1.2120$ x 10$^{-2}$ pc, r$_{\mathrm{in}} = 1.2120$ x 10$^{-3}$ pc, and dt$_{\mathrm{soft}} = 27.948$ years for our simulations. We also set r$_{\mathrm{bin}}= 100$ au for the radius under which we use slow-down algorithmic regularization.

\subsection{Hydrodynamics} \label{sec:hydro}  
We use the smoothed particles hydrodynamics (SPH) code \textsc{Gadget-2}~\citep{Springel2005} with a particle mass of 0.01 M$_{\odot}$ leading to 6.86$\times$10$^5$ particles in total. We convert the gas from \textsc{Flash} to \textsc{Gadget-2} using the method described in~\citet{Karam2024}. The gravitational force from the gas on the stars and vice-versa is calculated with \textsc{BHT}ree~\citep[based on][as implemented by Jun Makino]{Barnes1986} and the \textsc{Bridge}~\citep{Fujii2007} scheme. We use a bridge timestep of $2^5 \, \mathrm{dt}_{\mathrm{soft}} = 894.336$ years for our simulations.  

Our simulations do not include stellar feedback, stellar evolution, star formation or magnetic fields, in contrast with the \textsc{Torch} simulation from which we draw our initial conditions. It is however important to note that all stars in our simulations are younger than 3 Myr by the end of the runs, and that none of the stars would therefore have exploded in a supernova.

\subsection{Initial Conditions} \label{sec:ICs}

\begin{table}
\hspace{-1cm}
    \centering
    \begin{tabular}{cccc}
        \hline
        Run Name & Binaries & Prescription & Initial conditions \\
        \hline
        M0 & No & - & \textsc{Torch}\\
        M1 & No & - & \textsc{Torch}\\
        M2 & No & - & \textsc{Torch}\\
        AB1 & Yes & All binaries & \textsc{Torch}\\
        AB2 & Yes & All binaries & \textsc{Torch}\\
        AB3 & Yes & All binaries & \textsc{Torch}\\
        IB1 & Yes & Inner binaries & \textsc{Torch}\\
        IB2 & Yes & Inner binaries & \textsc{Torch}\\
        IB3 & Yes & Inner binaries & \textsc{Torch}\\
        WB1 & Yes & Wide binaries & \textsc{Torch}\\
        WB2 & Yes & Wide binaries & \textsc{Torch}\\
        WB3 & Yes & Wide binaries & \textsc{Torch}\\
        PM & Yes & Inner binaries & Plummer merger\\
        PB & Yes & Inner binaries & Isolated Plummer\\
        \hline
        
    \end{tabular}
    \caption{Overview of our simulations. Columns: the name of each simulation,  whether the simulation has a primordial binary prescription, the prescription used for primordial binaries, and the initial conditions for the gas and cluster shape (see text for more details). Different numbers in the run names correspond to different random seeds.}
    \label{tab:runs}
\end{table}

We summarize the different sets of initial conditions in Table~\ref{tab:runs} and in the text below.
For our fiducial initial conditions (M0), we take the stars from the \textsc{Torch} simulation run at 2.13 Myr. The distribution of the stars is shown in Figure~\ref{fig:Torch ICs}. 
The fiducial initial conditions have 3047 stars for a total mass of 1923.8 M$_{\odot}$. The stars are split into two clusters using a simple positional argument.
For the runs in which we introduce binaries, each system (i.e. single star or binary) is placed so that its centre of mass  is at the position of the star from M0 that has the mass closest to the system mass. This preserves the shape of the clusters, which is independent of the presence of binaries for embedded clusters~\citep[see][]{Cournoyer-Cloutier2023}. We also include two other runs without primordial binaries, \textsc{M1} and \textsc{M2}, where we randomly changed the mass of each star slightly from its mass in the \textsc{Torch} run but preserved each cluster's total mass.

We use three models for binaries, based on the observations presented in~\citet[][for stars above $\sim$ 1 M$_{\odot}$]{Moe2017} and~\citet[][for M-dwarfs]{Winters2019}. In the \textit{all binaries} models (AB1, AB2, AB3), we sample the full distribution of companions in the Galactic field using the technique presented in~\citet{Cournoyer-Cloutier2021}. In the \textit{inner binaries} models (IB1, IB2, IB3), we sample a distribution of inner companions; for intermediate- and high-mass stars, with a large triple fractions, this shifts the distribution to shorter periods. For each model, we vary the random seed to get the three different samplings. In the \textit{wide binaries} models (WB1, WB2, WB3), we impose a minimum period cut-off, following~\citet{Ramirez-Tannus2021},
\begin{equation}\label{eq:RT}
    P_{min} = 10^{(5 \pm 1)} \,  \mathrm{exp}\Big({-t/0.19^{+0.06}_{-0.04}}\Big) + 1.40
\end{equation}
where $P_{min}$ is the minimum period in days and $t$ is the cluster age in Myr. They derive this minimum period from the observed velocity dispersion of OB stars in young star clusters, which follows
\begin{equation}
    \sigma_{1\mathrm{D}} = (13.3 \pm 1.1) \, t + (2.0 \pm 0.1)
\end{equation}
We pick three possible values for the age of the cluster: 0.66 Myr (for the formation time of the first star), 0.21 Myr (for the average age of the simulated stars at the start of our zoom-in simulations) and 0 Myr (since the mean formation time of the stars in the full simulation is slightly larger than 2.13 Myr). Using those ages, we modify the IB1 initial conditions, so that all systems made up of two OB stars (i.e. all systems where both stars have a mass of at least 2 M$_{\odot}$) have an orbital period of at least P$_{\mathrm{min}}$. We report the minimum period for an OB binary at the start of each simulation in Table~\ref{tab:runaways}, as P$_{\mathrm{OB, min}}^0$. Although M0, M1 and M2 do not include any primordial binaries, they do include binaries formed through dynamics in the original \textsc{Torch} simulation. Those binaries have longer orbital periods than primordial binaries~\citep[see e.g][for a comparison between the properties of primordial and dynamically-formed binaries]{Cournoyer-Cloutier2021}.

\begin{figure}[tb!]
    \centering
    \includegraphics[width=0.95\linewidth, clip=True, trim=0cm 0cm 0cm 0cm]{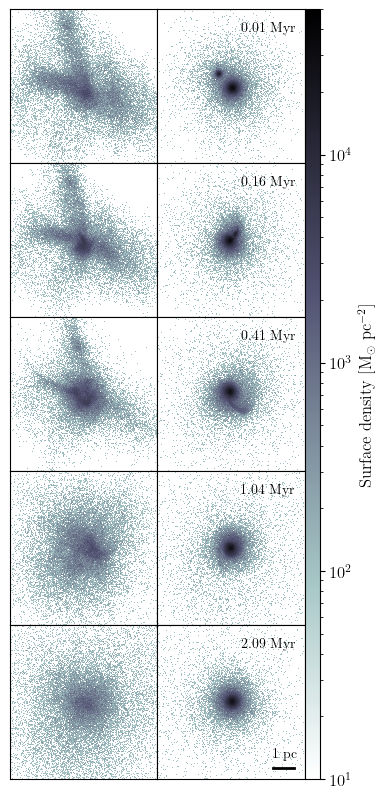}
    \caption{Gas surface density for the merger inherited from the \textsc{Torch} simulation (left) and for the merger of two isolated Plummer spheres (right). The maximum surface density in the Plummer spheres simulation is about one order of magnitude higher than in the merger from the \textsc{Torch} simulation.}
    \label{fig:Torch Plummer merger}
\end{figure}

To investigate the effects of the background gas on the binaries~\citep[as it was shown to be important for
the merger itself by][]{Karam2024}, we also simulate an idealized merger and an isolated cluster of the same mass, both without background gas. We simulate a merger of two Plummer spheres of gas and stars, with the same stellar and gas masses as the merging clusters (PM). The relative velocity and impact parameter are also kept fixed, as well as the total gas mass, but the gas and stars are initialized as pairs of relaxed~\citet{Plummer1911} spheres with densities consistent with young massive clusters, following~\citet{Karam2022}. We also simulate an isolated Plummer sphere with the total gas and stellar mass (PB). Both simulations use exactly the same stars and binaries as the IB1 simulation. 
A visual comparison of the gas between the merger inherited from the full initial conditions and the merger of two Plummer spheres is shown in Figure~\ref{fig:Torch Plummer merger}.

\section{Results}\label{sec:results} 

\subsection{Binary disruption}

\begin{figure}[tb!]
    \centering
    \includegraphics[width=\linewidth, clip=True, trim=0cm 0cm 0cm 0cm]{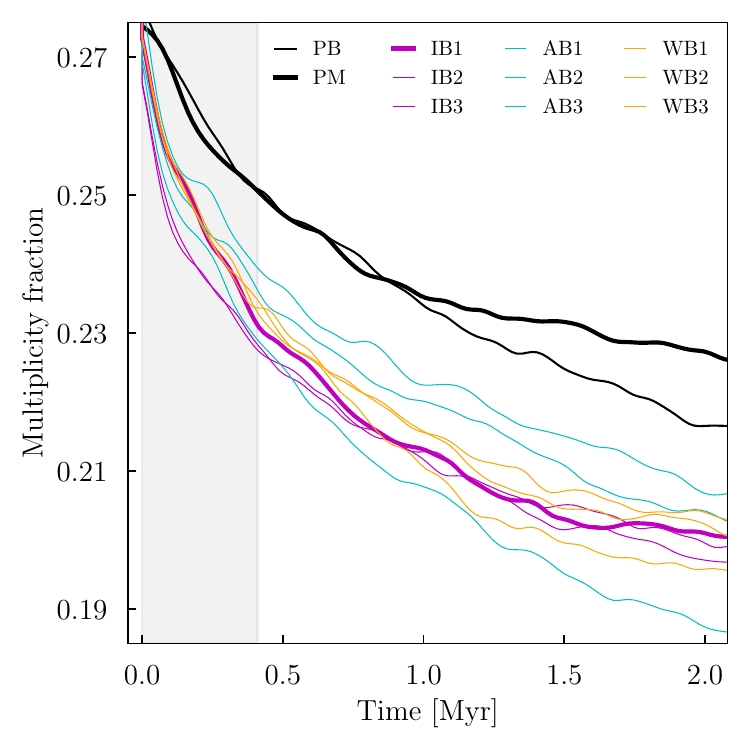}
    \caption{Binary fraction as defined in Equation~\ref{eq:bfrac} as a function of time for runs with primordial binaries. In merger runs with \textsc{Torch} initial conditions, about half of the total decrease in binary fraction takes place during the merger, which ends at 0.41 Myr.}
    \label{fig:fraction}
\end{figure}

The binary fraction as a function of time is shown in the upper panel of Figure~\ref{fig:fraction} for simulations with primordial binaries.
We calculate the binary fraction as
\begin{equation}\label{eq:bfrac}
    \mathcal{F}_{b} = \frac{B}{S + B}
\end{equation}
where $B$ is the number of stars with a bound companion within 10,000 au and $S$ is the number of stars without a companion within 10,000 au. Binaries wider than 10,000 au are not included in our statistics as they are very short-lived in dense stellar environments. 

The initial binary fraction is $\sim$ 27\% in all simulations with binaries. In the merger simulations with initial conditions inherited from the full \textsc{Torch} simulation (all the IB, AB and WB runs), the final binary fraction is between $\sim$ 19\% and $\sim$ 21\%, and the binary fraction at the end of the merger ($\sim$ 0.40 Myr) is around 23-24\%. For each of these simulations, about half of the decrease in the binary fraction takes place during the merger. The binary fraction therefore decreases by the same amount over the first 0.40 Myr and the last 1.60 Myr of the simulation, which indicates more rapid changes during the merger itself. The largest decrease takes place during the first $\sim$ 0.1 Myr of the simulation for all simulations realized with those initial conditions. We also note that the WB runs, which have all the same stars as IB1 and all the same binary systems for primaries less massive than 6 M$_{\odot}$, follow almost exactly the same decrease in binary fraction as IB1 during the merger. When the sampling for the binaries is changed, some scatter is introduced but the overall trend remains. The AB runs show the largest amount of scatter. Since the binaries are sampled from the parameter space of all companions rather than inner companions only, there is a larger range in the distributions of semi-major axes and binding energies of the binaries, leading to more scatter in the fraction of disrupted systems.

In the Plummer sphere merger (PM), the binary fraction decreases from $\sim$ 27\% to $\sim$ 23\% over the course of the simulation, with a steeper decrease in the first 0.9 Myr. This is the simulation that has the highest binary fraction at the end of the simulation. The binary fraction in the isolated Plummer sphere (PB) decreases from $\sim$ 27\% to $\sim$ 22\%. The binary fractions in PM and PB follow the same decrease in the first 0.9 Myr of the simulation, but diverge in the second half of the simulation. 

\begin{figure}[tb!]
    \centering
    \includegraphics[width=\linewidth]{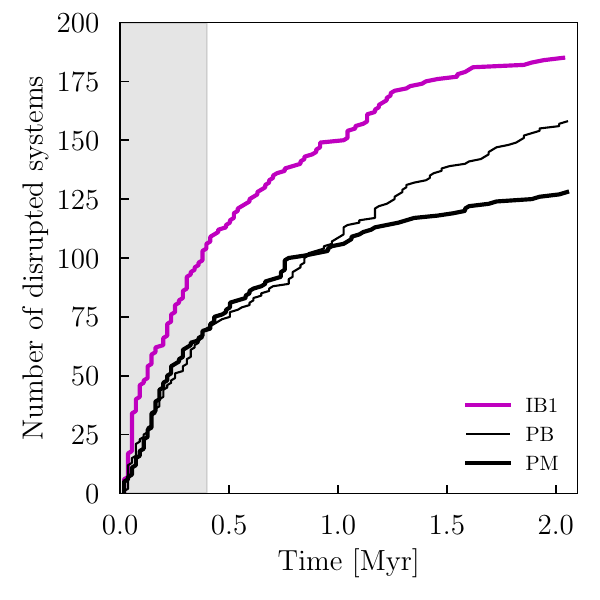}
    \caption{Cumulative number of disrupted systems for IB1, PB and PM. The three simulations have the same initial population of binaries.}
    \label{fig:disruption} 
\end{figure}

\begin{figure}[tb!]
    \centering
    \includegraphics[width=\linewidth]{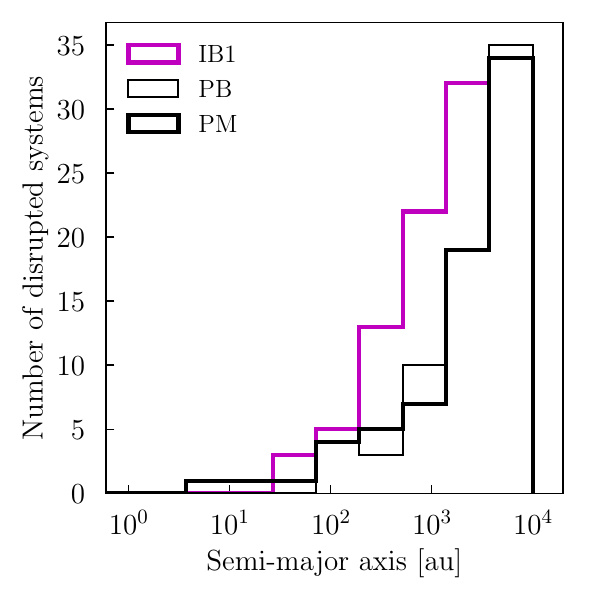}
    \caption{Semi-major axes of binary systems disrupted in IB1, PB and PM during the first 0.41 Myr of the simulation, which corresponds to the time required for the merger to complete in IB1. The three simulations have the same initial population of binaries.}
    \label{fig:a disrupted} 
\end{figure}

We plot the cumulative number of disrupted binaries as a function of time for all runs that use the IB prescription for primordial binaries, and present it in  Figure~\ref{fig:disruption}. For merger simulations, most of the disruption takes place during the merger. It continues throughout the entire merger, but peaks in the early stages of the merger, during the first $\sim$ 0.1 Myr. We see a similar behaviour in runs with the other two binary prescriptions. More than half of the total disruptions take place during the merger, with roughly 100 binary systems being disrupted in the first 0.4 Myr. We also note that the distribution of disruptions as a function of time is almost identical for the IB runs during the merger, while they diverge at later times. Similar behaviour is observed in the AB and WB runs. 
The PM and PB runs also show disruption at early times, although not to the extent observed in the IB, AB, and WB runs.  

All runs with primordial binaries and gas initial conditions inherited from a GMC-scale simulation show an excess of disruption during the merger. This excess is caused by the disruption of systems with smaller semi-major axes during the merger, as shown in Figure~\ref{fig:a disrupted}. We compare the distributions of semi-major axes for the binaries disrupted within the first 0.41 Myr of the simulation in IB1, PB and PM using a two-sample Kolomogorov-Smirnov (KS) test. We find that the semi-major axes of disrupted systems for PB and PM are consistent with being drawn from the same distributions. We are however confident at respectively 98.3\% and 98.8\% that the semi-major axes of the disrupted systems in IB1 are smaller than those in PB and PM. 

\subsection{Unbound stars}

\begin{figure}[tb!]
    \centering
    \includegraphics[width=\linewidth, clip=True, trim=0cm 0cm 0cm 0cm]{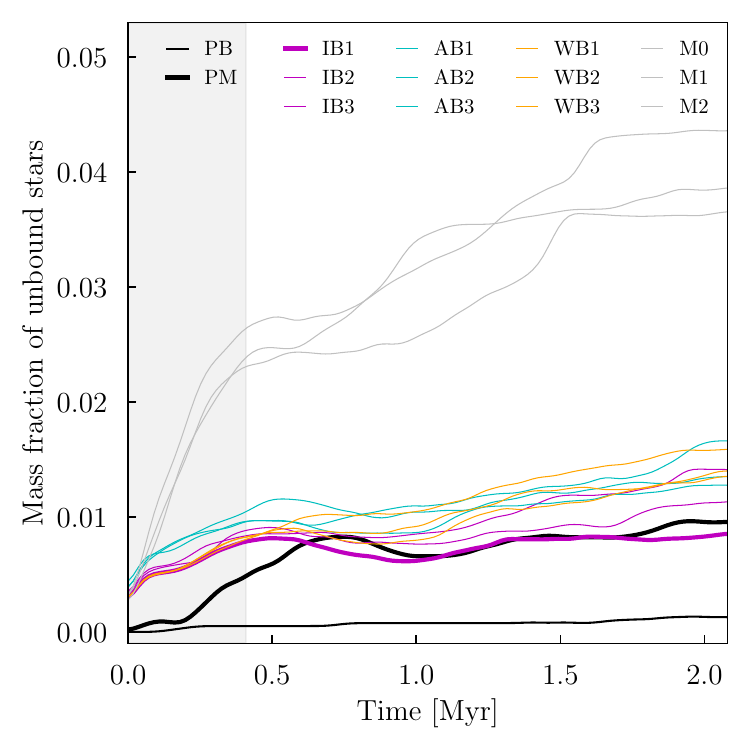}
    \caption{Mass fraction of unbound stars as a function of time, for all simulations.}
    \label{fig:UnboundFrac}
\end{figure}

We present the unbound mass fraction for stars for all simulations in Figure~\ref{fig:UnboundFrac}. 
The fraction of unbound stars increases during the merger for all merger simulations~\citep[as expected from][]{Karam2022, Karam2024}, while almost all stars remain bound to their host cluster in PB (the isolated Plummer sphere). A smaller fraction of stellar mass becomes unbound at early times in \textsc{PM} than in the other simulations with a primordial binary prescription, which behave very similarly to one another. By the time the merger ends in \textsc{PM}, at 0.60 Myr, the mass fraction of unbound stars has reached the same value as in the other simulations with primordial binaries. The most important difference is between the runs with primordial binaries, and those without primordial binaries. M0, M1 and M2 all have a similar mass fraction of unbound stars at the end of the merger, which is about 2.5 times higher than the mass fraction of unbound stars in the runs with \textsc{Torch} initial conditions and primordial binaries. 
This suggests that the underlying physical process responsible for the production of unbound stars during sub-clusters mergers is the same that leads to the disruption of binaries. In the absence of primordial binaries, the energy from the merger is converted to kinetic energy of the stars, leading to an increase in the fraction of unbound stars. In the presence of primordial binaries, however,  the energy from the merger is used to disrupt binaries and is not sufficient to also unbind stars from the cluster.

\begin{figure}[tb!]
    \centering
    \includegraphics[width=\linewidth, clip=True, trim=0cm 0cm 0cm 0cm]{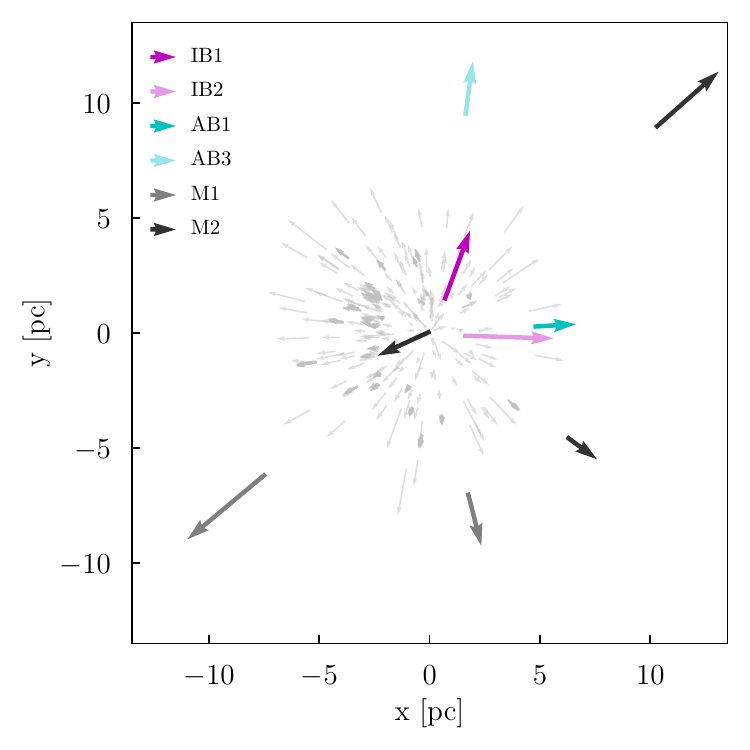}
    \caption{Projected direction of motion in the xy plane of all unbound stars in simulations with \textsc{Torch} initial conditions, at the end of the merger. The length of the arrow is proportional to the projected velocity and the origin of the arrow corresponds to the location of the star. Runaway stars are color-coded by simulation while stars with $v_r <$ 30 km/s are shown in light grey.
    }
    \label{fig:Quiver}
\end{figure}

\begin{figure}[tb!]
    \centering
    \includegraphics[width=\linewidth, clip=True, trim=0cm 1.5cm 0cm 1.5cm]{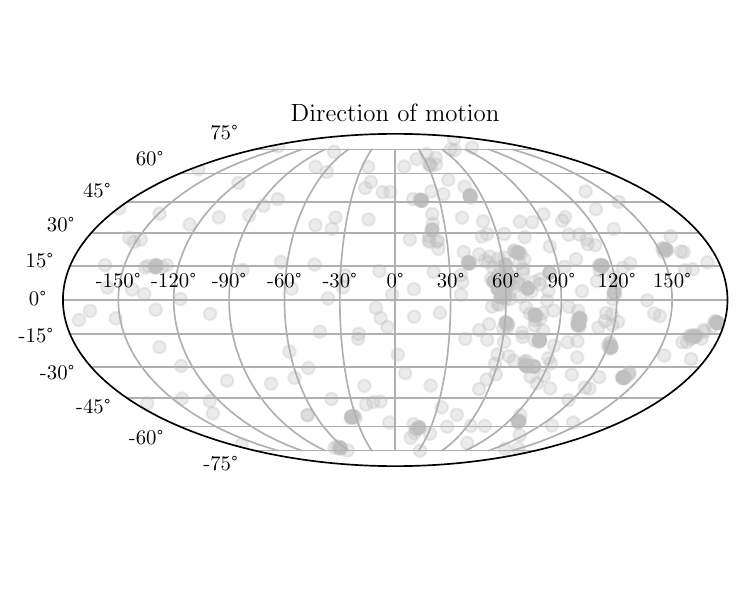}
    \caption{Mollweide projection of the direction of motion of all unbound stars in simulations with \textsc{Torch} initial conditions, at the end of the merger.
    }
    \label{fig:Mollweide}
\end{figure}

For each unbound star, we calculate its three-dimensional radial velocity $v_r$ relative to the cluster's center of mass. The distributions of velocities in the simulations with and without primordial binary distributions are similar. We label unbound stars with $v_r \geq 30$ km/s as runaways. We report the number of runaways N$_{\mathrm{runaways}}$ in Table~\ref{tab:runaways}, along with the number of runaways more massive than 6~M$_{\odot}$ (N$_{\mathrm{runaways, M}}$), the number of unbound stars more massive than 6~M$_{\odot}$ (N$_{\mathrm{unbound, M}}$) and the number of runaways initially in binaries (N$_{\mathrm{runaways,} b}$). The only merger simulation that does not produce any runaway stars is \textsc{AB2}, which however has two unbound stars with $v_r \geq 25$ km/s, just below our runaway cut-off. PB, on the other hand, does not produce any runways, and the maximum radial velocity for an unbound star is 12.8 km/s. 

The runaways produced during the merger for simulations with \textsc{Torch} initial conditions are shown in Figure~\ref{fig:Quiver}, along with the unbound stars with $v_r \leq 30$ km/s. We also calculate the direction of motion of all unbound stars and the runaways stars after the mergers. We show the positions and directions of motion of all runaways produced during the merger in Figure~\ref{fig:Quiver}. There are no runaways moving towards the (-x, +y) quadrant, which is the original location of the less massive cluster. If we consider all unbound stars produced during the merger for all \textsc{Torch} runs, however, the distribution of directions of motions peaks in the -x direction. Comparing the distribution of angles to uniform distributions in $\phi$ and $\theta$ with a KS test, we are confident at $>$ 99.9\% that the direction of motion of the unbound stars is not uniform. We present the three-dimensional direction of motion of unbound stars in Figure~\ref{fig:Mollweide}. The directions of motions are clustered around an angle $(\phi, \theta) = (60^{\mathrm{o}}, 0^{\mathrm{o}})$.

\begin{table*}[tb!]
    \centering
    \begin{tabular}{cccccccccccccc}
        \hline
        Name & N$_{\mathrm{runaways}}$ & N$_{\mathrm{runaways, M}}$ & N$_{\mathrm{unbound, M}}$ & N$_{\mathrm{runaways, }b}$ & & $\mathcal{F}_{b \mathrm{, i}}$ & & $\mathcal{F}_{b, m}$ & & $\mathcal{F}_{b, f}$ & & P$_{\mathrm{OB, min}}^{\mathrm{0}}$ [d] & P$_{\mathrm{OB, min}}^{\mathrm{f}}$ [d]\\
        \hline
        M0 & 2 & 0 & 4 & 0 & & 0.10  & & 0.06 & &  0.00 & & 2.68 x 10$^6$ & 7.42 x 10$^3$\\
        M1 & 10 & 0 & 2 & 0 & & 0.08 & & 0.05  && 0.00 & & 2.16 x 10$^6$ & 1.31 x 10$^4$\\
        M2 & 9 & 1 & 5 & 0 & & 0.11 & & 0.09 & & 0.00 & & 1.28 x 10$^7$ & 7.82 x 10$^3$\\
        IB1 & 2 & 0 & 1 & 1 & & 0.26 & & 0.23 & & 0.10 & & 48.2 & 48.2\\
        IB2 & 5 & 0 & 2 & 4 & & 0.31 & & 0.26 & & 0.13 & & 21.0 & 21.0\\
        IB3 & 8 & 2 & 2 & 5 & & 0.33 & & 0.26 & & 0.11 & & 39.5 & 32.9\\
        AB1 & 3 & 0 & 1 & 2 & & 0.35 & & 0.31 & & 0.17 & & 106 & 105\\
        AB2 & 0 & 0 & 0 & 0 & & 0.31 & & 0.25 & & 0.09 & & 40.9 & 40.7\\
        AB3 & 2 & 0 & 1 & 0 & & 0.37 & & 0.30 & & 0.13 & & 40.2 & 34.7\\
        WB1 & 7 & 0 & 0 & 1 & & 0.33 & & 0.26 & & 0.04 & & 3.10 x 10$^3$ & 1.44 x 10$^3$\\
        WB2 & 4 & 1 & 2 & 2 & & 0.35 & & 0.38 & & 0.08 & & 3.31 x 10$^4$  & 4.46 x 10$^3$\\
        WB3 & 3 & 1 & 4 & 3 & & 0.26 & & 0.24 & & 0.07 & & 1.00 x 10$^5$  & 3.95 x 10$^3$\\\
        PB & 0 & 0 & 0 & 0 & & 0.50 & & 0.25 & & 0.00 & & 48.2 & 48.2\\
        PM & 6 & 0 & 1 & 2 & & 0.29 & & 0.16 & & 0.13 & & 48.2 & 48.2\\
        \hline
    \end{tabular}
    \caption{Properties of unbound stars and OB binaries in the different simulations. Columns: run name, number of runaways, number of runways with mass above 6~M$_{\odot}$, number of unbound stars with mass above 6~M$_{\odot}$, number of runaways originally in binaries, fraction $\mathcal{F}_{b \mathrm{, i}}$ of stars unbound at the end of the simulation originally in binaries, fraction $\mathcal{F}_{b \mathrm{, m}}$ of stars unbound at the end of the simulation in binaries at the end of the merger, fraction $\mathcal{F}_{b \mathrm{, f}}$ of stars unbound at the end of the simulation in binaries at the end of the simulation, minimum initial period for an OB binary, minimum period for an OB binary at the end of the simulation.}
    \label{tab:runaways}
\end{table*}

We also report the binary fraction for stars that are unbound at the end of the simulation in Table~\ref{tab:runaways}. We calculate the binary fraction at the start of the simulation, at the end of the merger, and at the end of the simulation. The binary fraction for those stars decreases in all simulations, consistent with our expectation of disrupted binaries leading to unbound stars. We note that for simulations without primordial binaries, which only include wide, dynamically-formed bianries, the binary fraction of unbound stars at the end of the simulation is 0, while it ranges from 7 to 17\% for all runs with primordial binaries (except PB). 

\subsection{Outcomes for massive binaries}

We now consider specifically how massive OB binaries evolve throughout the simulations. We consider OB binaries to be systems with a primary mass above 6 M$_{\odot}$ (the lower mass threshold considered by~\citealt{Ramirez-Tannus2021}) and companion mass above 2 M$_{\odot}$ (the lower mass limit for B-type main sequence stars). We are interested in two main features: can cluster mergers with background gas harden massive binaries, and can they disrupt them?

We first turn our attention to the OB binary with the shortest orbital period in each simulation. The minimum orbital period for an OB binary at the beginning and at the end of each simulation is reported in Table~\ref{tab:runaways}. There are no significant changes to the orbital period of the OB star with the shortest period in IB1, IB2, AB1, AB2, PB or PM. In IB3, the OB binary with the shortest period at the end of the simulation originally had a an orbital period of 123 days, and was hardened to an orbital period of 32.9 days through an exchange at 1.92 Myr. In AB3, the OB binary that had initially had the shortest orbital period, at 40.2 days, got hardened to an orbital period of 34.7 days, likely through a few-body encounter. 

In simulations with initially wide massive binaries (WB1, WB2 and WB3), the minimum orbital period for OB binaries decreases significantly. In WB1, it decreases by a factor of $>$ 2, going from $\sim$ 3100 days to $\sim$ 1440 days. The OB binary that has the shortest period at the end of the simulation however initially had an orbital period slightly longer than 13,000 days, and was hardened through an exchange at 1.23 Myr followed by a few-body interaction at 1.59 Myr. In WB2 and WB3, the OB binary that has the shortest orbital period at the end of the simulation also had the shortest possible period at the beginning of the simulation. Those periods were shorted by a factor of 7.4 and 25 in WB2 and WB3 respectively; we also note that the shortest period is shorter in WB3 than WB2 at the end of the simulation.
This hardening is the expected behaviour 
from observations of OB binaries in young clusters~\citep[][orbital periods decrease during cluster formation]{Ramirez-Tannus2021}. It however must be noted than none of the OB binaries in WB runs reaches an orbital period below 1000 days. 
Additionally, exchanges and interactions that significantly shorten the orbital period of OB binaries all take place well after the merger.  

We look for disrupted OB binaries in all simulations that include primordial binaries. In the runs with \textsc{Torch} initial conditions, all the disruption of OB binaries takes place after the merger. There is one early disruption in PM. We stress that energetic binaries can be disrupted early if they encounter other energetic binaries. If we calculate the hard-soft limit for the resultant clusters, we find values around 2 x 10$^{44}$ erg, with a maximum value of 3.10 x 10$^{44}$ erg. Almost all the disrupted OB binaries were more energetic than this limit: this suggests that the short-period OB binaries observed in young star clusters might not be fully representative of the primordial population, despite their high binding energies.
We also note that OB binaries were disrupted in all merger runs with primordial binaries, but not in the isolated Plummer sphere. 

Only one massive star becomes unbound from its host cluster the during the merger: a star with mass 68 M$_{\odot}$ in \textsc{M0}, the fiducial set of initial conditions. This star is the second most massive in the simulation. It forms a binary with semi-major axis $\sim$ 300 au with a star with mass 63 M$_{\odot}$ during the merger. It is then disrupted by an encounter with the most massive star in the simulation, which has a mass of 85 M$_{\odot}$; after the encounter, the 85 M$_{\odot}$ and 63 M$_{\odot}$ stars form a binary with semi-major axis $\sim$ 80 au. \textsc{M0}, \textsc{M1} and \textsc{M2} produce the most massive unbound stars: by the end of the simulation, \textsc{M0} has unbound stars with masses 22, 23, 50 and 68 M$_{\odot}$, \textsc{M1} has unbound stars with masses 49 and 66 M$_{\odot}$, and \textsc{M2} has an unbound stars with mass 29 M$_{\odot}$. They have all been identified as members of a wide binary in at least one snapshot, but those did not persist between snapshots, indicating that a few-body interaction took place. 
This illustrates how wide, dynamically-formed binaries can results in the ejection of massive stars or large numbers of ejected stars, due to their larger dynamically cross-sections. Indeed, the other two simulations without primordial binaries, \textsc{M1} and \textsc{M2}, also have the largest number of runaway stars produced during the merger, and during the full simulation. 
On the other hand, the mass of the most massive unbound star in any simulation with primordial binaries is $\sim$ 11 M$_{\odot}$.

\section{Discussion}\label{sec:discussion}

\subsection{Causes of binary disruption during hierarchical cluster assembly}

There are several differences between the simulations using the initial conditions from \textsc{Torch} and those using idealized Plummer models. We discuss the effects of the stellar and gas distributions below. 

The stellar density profiles for IB1, PB, and PM are very similar. All three simulations reach densities $\gtrsim$ 10$^4$ M$_{\odot}$ pc$^{-3}$  (and therefore number densities $>$ 10$^4$ pc$^{-3}$) in their central regions. If high stellar densities were the cause of the excess disruption we observe in the \textsc{Torch} runs, we would expect the mass or number density to be significantly higher in the \textsc{Torch} runs than in the idealized models, which is not the case. 
The gas distribution however differs strongly between the \textsc{Torch} runs -- which were extracted from a GMC simulation -- and the idealized Plummer models. The central gas density is about an order of magnitude higher in the Plummer models than in the \textsc{Torch} models, as illustrated in Figure~\ref{fig:Torch Plummer merger}. Conversely, the gas distribution is more extended in the \textsc{Torch} models, and includes background gas. 

The more realistic gas distribution contributes in two ways to the excess disruption of binaries observed in the \textsc{Torch} models. First, the presence of background gas promotes the merger between the sub-clusters, as found by~\citet{Karam2024}. This means that the stars from the smaller cluster are accreted unto the larger cluster much more quickly in the \textsc{Torch} simulations than in the Plummer spheres' merger. This rapid accretion means that binaries in either cluster are more likely to encounter other binaries and stars, and therefore have a larger chance to be disrupted. On the other hand, the Plummer spheres do not immediately merge, and the two clusters can be easily identified after the first passage. Second, the potential is shallower in the \textsc{Torch} runs, due to the more extended gas distribution. This allows binaries with higher binding energies (and therefore smaller semi-major axes) to be disrupted by allowing the stars to move apart more easily after an encounter. We note that the potential increases sharply during the first $\sim$ 0.2 Myr of the simulations, during which the disruption rate of binaries peaks in all \textsc{Torch} runs.

\subsection{Unbound stars in embedded star clusters}

Recent observations have found a possible anisotropy in the distribution of runaway stars around the young massive star cluster R136~(Stoop et al. subm.). Using simulations of massive cluster formation, \citet{Polak2024} attribute this effect to the production of runaways from hierarchical cluster assembly inside a GMC. They find that an entire sub-cluster can become a system of runaway stars, due to the large tidal forces involved in the merger which completely destroys an incoming cluster. Our simulations however probe a lower cluster mass range, resulting in a smaller number of runaways, which is not sufficient to investigate whether those stars have a preferred direction. Considering the full population of unbound stars -- of which runaways are only a subset -- however allows us to conclude that stars ejected during sub-cluster mergers inside GMCs have a preferred direction of motion, even in a lower-mass regime. 

\subsection{Implications for massive cluster formation}
Our results offer several insights into the process of (massive) star cluster formation. First, the disruption of binaries during hierarchical cluster assembly within giant molecular clouds offers a natural explanation for the lower wide binary fraction found in the ONC than in other young star-forming regions~\cite[e.g.][]{Duchene2018}. As the ONC has been proposed to be the recent site of a merger between gas-rich sub-clusters~\citep[see e.g.][]{Fujii2022}, the disruption of the wide binaries during the merger process would be consistent with our findings. 
\citet{Howard2018} found that massive clusters acquire about half of their stellar mass via gas-rich mergers. 
Such a mode of cluster assembly could contribute to the low binary fractions observed in globular clusters~\citep[e.g.][]{Milone2016}. 

Binary disruption during gas-rich or gas-driven sub-cluster mergers leading to a shift towards smaller semi-major axes could also explain the apparent discrepancy between the results of~\citet{Cournoyer-Cloutier2021} and ~\citet{Parker2014} or~\citet{Torniamenti2021} regarding the evolution of field-like population of binaries in star clusters. In~\citet{Cournoyer-Cloutier2021}, we found that a field-like distribution of semi-major axes, like the one used in the AB runs, shifts to smaller values during hierarchical star cluster formation. 
On the other hand, for a similar distribution of binaries, \citet{Parker2014} use fractal initial conditions for their stars, while~\citet{Torniamenti2021} use stellar positions inherited from the gas distribution. Neither, however, included gas along with the N-body dynamics while investigating the binaries. Both found the distribution of semi-major axes statistically unchanged by the process of (gas-free) cluster assembly. Along with the results presented in this paper, this suggests that changes in the gas potential that coincide with sub-clusters merger play a critical role in disrupting binaries during cluster formation. 

Another important implication of our results concerns the hardening of massive binary systems. The density in the inner 0.3 pc of IB1 is roughly 1.2 x 10$^{4}$ M$_{\odot}$ pc$^{-3}$ at the start of the simulation, and 8.6 x 10$^{3}$ M$_{\odot}$ pc$^{-3}$ after 0.9 Myr, around the typical central densities for young massive clusters~\citep{PortegiesZwart2010}. Despite those high densities, however, simulations that started without any close massive binaries did not succeed in forming any OB binary with an orbital period shorter than 1000 days, while the most conservative choice proposed by~\citet{Ramirez-Tannus2021} in Equation~\ref{eq:RT}, using a cluster age of 2 Myr, gives a period of 337 days, about a factor of 4 shorter than the shortest period OB binary we get in WB1. Although our simulated clusters explore a lower mass range than the clusters studied by~\citet{Ramirez-Tannus2021}, our results suggest that the orbital parameters of OB binaries can be modified on timescales shorter than the cluster formation timescale.

\section{Summary}\label{sec:summary}
We have conducted simulations of stellar sub-cluster mergers with different realistic prescriptions for the initial distribution of binaries. The shapes and masses of the sub-clusters, as well as the background gas, were taken from larger-scale simulations of cluster formation within a giant molecular cloud. We find that massive binaries can be disrupted or undergo significant changes to their orbital periods over timescales shorter than the cluster-formation timescale, in all of our merger simulations. The observed distributions of OB binaries is likely not the same as their formation distribution even in very young clusters and associations. 

We also find that the merger process -- and by extension, hierarchical cluster formation -- lowers the binary fraction and disrupts a large number of binaries during the merger, in excess of what is observed for an isolated cluster or an idealized model of cluster merger. The disrupted systems also tend to have smaller semi-major axes than in idealized models. This excess disruption is attributed to the rapid merger driven by the background gas distribution, as well as the shallower potential. Sub-cluster mergers result in stars becoming unbound from their host cluster, with a larger mass fraction of unbound stars when primordial binaries are taken into account. Simulations without primordial binaries, in which the most massive stars tend to pair up in a wide, dynamically-formed binaries, are capable of ejection higher-mass stars for the cluster mass and density regime we consider. We further note that stars that become unbound during the merger tend to move in the same direction. 

We conclude that the production of unbound stars with a preferred direction of motion is a natural consequence of sub-cluster mergers within a GMC, extending the results of~\citet{Polak2024} to lower cluster masses.
We further argue that viewing binary disruption as a by-product of sub-cluster mergers within a giant molecular cloud offers a natural explanation for the lower binary fraction observed in denser star-forming environments and in globular clusters, which form hierarchically through subsequent mergers.

\begin{acknowledgments}
CCC is grateful for the hospitality of Leiden University, where this work was started during a visit in June 2023. The authors thank the referee for comments that improved the clarity of the paper. The authors warmly thank Eric Andersson, Sabrina Appel, Mordecai-Mark Mac Low, Stephen McMillan and Brooke Polak for ongoing discussions about \textsc{Torch}. The authors also thank Veronika Dornan, William Harris and Marta Reina-Campos for helpful discussions. CCC and JK are grateful to Gwendolyn Eadie for very valuable insights regarding the choice of initial stellar distributions. 

CCC is supported by a Canada Graduate Scholarship -- Doctoral (CGS D) from the Natural Sciences and Engineering Research Council of Canada (NSERC).  The visit to Leiden University was made possible by a Michael Smith Foreign Study Supplement (CGS MSFSS) held by CCC at the Max Planck Institute for Astrophysics in Summer 2023. JK and AS are supported by NSERC. This research was enabled in part by support provided by Compute Ontario (\url{https://www.computeontario.ca/}) and the Digital Research Alliance of Canada (\url{alliancecan.ca}) via the research allocation FT \#2665: The Formation of Star Clusters in a Galactic Context. The \textsc{Torch} simulation used for initial conditions was run on Snellius through the Dutch National Supercomputing Center SURF grants 15220 and 2023/ENW/01498863.
\end{acknowledgments}

%


\software{
matplotlib~\citep{matplotlib}, numpy~\citep{numpy},
pynbody~\citep{pynbody},
yt~\citep{yt}} 





\bibliography{bibliography}{}
\bibliographystyle{aasjournal}



\end{document}